\documentclass[aps,prl,reprint,groupedaddress,nofootinbib]{revtex4-1}

\usepackage{amsmath}
\usepackage{amsfonts}
\usepackage{amssymb}
\usepackage{graphicx}
\usepackage{soul}
\usepackage{xcolor}

\begin{document}

%\preprint{}

\title{Testing Quantum Black Holes with Gravitational Waves}


\author{Valentino F.\ Foit}
\email[]{foit@nyu.edu}
%\thanks{}
\author{Matthew Kleban}
\email[]{kleban@nyu.edu}
\affiliation{Center for Cosmology and Particle Physics, Department of Physics, New York University}

\date{\today}

\begin{abstract}
We argue that  near-future detections of  gravitational waves from merging black hole binaries can test a long-standing proposal, originally due Bekenstein and Mukhanov, that the areas of black hole horizons are quantized in integer multiples of the Planck area times an $\mathcal O(1)$ dimensionless constant $\alpha$. This  condition quantizes the frequency of radiation that can be absorbed or emitted by a black hole.   If this quantization applies to the ``ring down'' gravitational radiation emitted immediately after a black hole merger, a single measurement consistent with the predictions of classical general relativity would rule out most or all (depending on the spin of the hole) of the extant proposals in the literature for the value of $\alpha$. A measurement of two such events for final black holes with substantially different spins  would rule out the proposal for any $\alpha$.  If the modification of general relativity is confined to the near-horizon region within the hole's light ring and does not affect the initial ring down signal, a detection of ``echoes'' with characteristic properties could still confirm the proposal.
\end{abstract}


\maketitle


\section{Introduction}

In classical general relativity, the event horizons of large black holes in vacuum are regions of low curvature, hardly distinguishable from flat spacetime to an inertial observer nearby.  However, the quantum mechanics of horizons has a number of surprising features, most notably that black holes radiate thermally at the Hawking temperature and that the entropy of the hole is proportional to its horizon area.  While for macroscopic black holes neither of these effects in itself is  observable in any realistic scenario, a number of authors have proposed that the quantum modifications go well beyond these subtle effects, such that the  physics of quantum black holes --- even very large, astrophysical holes --- is dramatically different from their classical counterparts.  

One of the most long-standing ideas is due to Bekenstein and Mukhanov \cite{Bekenstein1974,Bekenstein:1995ju}, who proposed that the area $A$ of black hole horizons is quantized in units of the Planck area:
\begin{align} \label{quant}
	A = \alpha l_{P}^{2} N = \alpha \hbar G N,
\end{align}
where $N$ is an integer, $\alpha$ is an $\mathcal{O}(1)$ dimensionless coefficient, and we have  set $c=1$.  One might naively expect that such a tiny quantum of area would have no observable implications for large black holes.  However, \eqref{quant} implies that the spectrum of emission or absorption of radiation by quantum black holes occurs in a series of almost exactly evenly-spaced lines. For a Schwarzschild (spinless) black hole $A = 4 \pi {r_s}^2 = 4 \pi (2 G M)^{2}$, so $\Delta A =  \alpha \hbar G  \Delta N =  32 \pi  G^{2} M \Delta M$, and hence 
\begin{align}\label{quant2}
	\omega_{n} = - \frac{\Delta M}{\hbar} = \frac{n \alpha}{32 \pi} \frac{1}{M G} = \frac{n \alpha}{16 \pi} \frac{1}{r_s}
\end{align}
Here $n = - \Delta N$ is the change in the area quantum.  One sees that if the Bekenstein-Mukhanov proposal is correct, black holes behave more like atoms than large, opaque objects \cite{Bekenstein:1997bt}.  In particular they cannot absorb any radiation with wavelength longer than the fundamental mode of \eqref{quant2} \cite{Mukhanov:1986me}.

Each level must have a degeneracy of order $e^{S}$, where $S = A/4 G \hbar$ is the entropy of the hole.  Ordinarily one  expects  degeneracy to be split due to interactions that break whatever symmetry or coincidence was responsible for it.  If so,  the quantization would be $\Delta M/M \sim e^{-S} $.  Since $S \sim 10^{78}$ for a solar mass black hole, this  would be totally unobservable for macroscopic holes.  Instead,   in the Bekenstein-Mukhanov proposal the width of each line is much smaller than the spacing between the lines, despite the huge degeneracy \cite{Mukhanov:1986me}. 
%\red{\st{In our opinion this is implausible.}}

In the next section, we will generalize equation \eqref{quant} to  black holes with spin and proceed to see how the resulting frequency quantization can be tested.
The wavelength of emitted and absorbed radiation is quantized in integer multiples of a fundamental wavelength \emph{of order the black hole horizon size}.  This is not necessarily a small effect!  For instance it is very noticeable in the spectrum of Hawking radiation, which, under this proposal, is very far from thermal blackbody. The spacing between the lines is such that if the sun's $\sim$ 6000 K  blackbody spectrum were replaced with the corresponding Bekenstein-Mukhanov spectrum for a black hole with 6000 K Hawking temperature, it would radiate in the optical only in a single, narrow band: with the popular choice $\alpha=4 \ln 3$ \cite{PhysRevLett.81.4293}, the sun would be a brilliant green. 

If the Bekenstein-Mukhanov proposal is correct, black holes are more similar to individual atoms with discrete line spectra than optically thick objects.  In fact, given the lack of a microphysical model that accounts for the horizon area quantization, perhaps the best analogy is to the Bohr model of the atom, rather than full quantum mechanics.   This heuristic nature of the proposal makes it difficult to analyze the theory in complex, dynamical situations such as mergers.  Nevertheless, we will argue that it can be tested, at least under certain assumptions.  

Our main result is that the frequency of the mode that dominates the radiation emitted by a perturbed classical black hole depends on the spin of the hole in a different way than the frequency determined by the area quantization rule.  Therefore, observations of the radiation emitted by perturbed black holes with differing values of spin can sharply distinguish between classical holes and the Bekenstein-Mukhanov proposal.

\subsection{The value of the parameter $\alpha$}

Various values have been suggested for the proportionality constant $\alpha$ in \eqref{quant}.  Bekenstein and Mukhanov \cite{Bekenstein:1995ju} proposed  $\alpha = 4 \ln q$, where $q$ is an integer. This follows if one requires that the number of states $e^{S}$ be an integer, since $e^S = e^{A/4G} = e^{N \alpha/4}$. Mukhanov  argued in particular for  $\alpha = 4 \ln 2 \approx 2.8$ \cite{Mukhanov:1986me}.   Hod advocated $\alpha = 4 \ln 3 \approx 4.4$ from an argument involving matching to the most highly-damped quasinormal mode (QNM) frequencies \cite{PhysRevLett.81.4293}. Arguments from ``canonical quantum gravity'' also indicate $\alpha = 4 \ln 3$ \cite{Dreyer:2002vy}. Maggiore prefers the substantially larger value $\alpha = 8 \pi \approx 25$, again by matching to highly-damped QNMs \cite{PhysRevLett.100.141301}, as do several analyses that extend Maggiore's methods to Kerr black holes \cite{Vagenas:2008yi, Medved:2008iq}. This was also the value originally suggested by Bekenstein in \cite{PhysRevD.7.2333}. Davidson \cite{Davidson:2011eu} suggests $\alpha = 8 \ln 2$ based on a ``holographic shell model" for black holes.

\section{Testing Bekenstein-Mukhanov with LIGO}

When a classical black hole is perturbed, it vibrates and equilibrates to a stationary configuration by emitting gravitational waves.
The late time behavior of this ring down can be described by a discrete set of exponentially decaying ``quasi-normal modes'' (QNMs).
Their spectrum can be determined by a linearized perturbation analysis around the black hole metric.  The  values of the real and imaginary parts of the QNM frequencies all scale  with the inverse radius $r \sim G M$ of the hole, since that is the only lengthscale in the problem, but they depend in a non-trivial way on the spin parameter of the hole $a$, the spherical multipole $l$, and a mode integer $k$. 

The allowed frequencies of emitted (or absorbed) gravitons under the Bekenstein-Mukhanov proposal for a spinning black hole of mass $M$ and angular momentum $J$ can be determined as follows. The change of the area is conjectured to be $\Delta A = - \alpha \hbar G n$,
%\begin{align}\label{deltaA1}
%	\Delta A = - \alpha \hbar G n,
%\end{align}
where again $n=-\Delta N$. The area of a Kerr black hole is $A = 8 \pi M^2 G^2 \left( 1 + \sqrt{ 1 - {J^2}/{G^2 M^4}} \right)$, so
$\Delta A ={\partial A}/{\partial M}|_J \Delta M + {\partial A}/{\partial J}|_M \Delta J$ and we find
\begin{align}\label{deltaA2}
	\Delta A = 16 \pi G^2 M \left( 1 + \frac{1}{\sqrt{1-a^2}}\right) \Delta M - \frac{8 \pi G a }{\sqrt{1-a^2}} \Delta J.
\end{align}
If the energy and angular momentum of the emitted/absorbed quanta are $\Delta M = -\hbar \omega_n$ and  $\Delta J = -\hbar m$,
%equating \eqref{deltaA1} and \eqref{deltaA2} 
%these $\Delta A$ 
this gives\footnote{This disagrees with \cite{Bekenstein:2015soa} by a factor of 4, which we believe is due to an error in that reference.}
\begin{align}\label{wfund}
	\omega_{n}  = (M G)^{-1} \frac{n \alpha \sqrt{1-a^2} + 8 \pi a m}{16\pi \left( 1 + \sqrt{1-a^2}\right)},
\end{align}
where $a\equiv J/GM^{2}$ is the dimensionless spin of the  hole ($0 \leq a \leq 1$). For a single graviton, $m = \pm 2$.\footnote{We will assume that only a single graviton is emitted in each quantum area transition. This is justified by the extremely weak interaction between two or more such gravitons that would be necessary for emission of more than one graviton \cite{Foit:2015wqa}, and by arguments of Bekenstein in \cite{Bekenstein:1997bt}.} The spectrum (\ref{wfund}) has the same scaling with $M G$ as the QNMs, but the dependence on $a$ is quite different (Figure \ref{fig:QNMvsBM1}).  

Presumably, quantization of the horizon area can be neglected and classical gravity should be trusted during the inspiral  phase of a binary merger when the holes are separated by distances much larger than their horizon radii.  During the last stages of the inspiral and just after the holes merge the physics is highly non-linear and it is unclear what the Bekenstein-Mukhanov proposal predicts.\footnote{Although whatever it does predict may well not agree with classical general relativity, and one could possibly derive even stronger constraints from that phase.}  However, once the holes merge the resulting object can be regarded as a larger black hole with mass $M < M_{1} + M_{2}$ and with an initial  anisotropic perturbation that then decays.  This decay is well-modeled as a linear perturbation to the Kerr metric that rings down exponentially in time.

As mentioned above, the emitted radiation has a frequency of order $1/G M$, which in order of magnitude is equal to that of quantum Hawking radiation.  That is, although the gravitational waves  observed by LIGO contain an enormous number of individual gravitons, each graviton was evidently produced by a quantum transition between energy levels separated by $\mathcal{O}(1)$ in units of the Hawking temperature.  If the Bekenstein-Mukhanov proposal correctly describes Hawking radiation, it must describe these levels.  Furthermore, during the ring down phase the perturbation is linear and only slightly modifies the area of the horizon.  Therefore it seems we may be justified in applying \eqref{wfund} to the radiation emitted during the ring down phase.

However, care must be taken at early times after the merger.
It is  known that QNMs do not necessarily describe the initial stages of the ring down.
This can be understood due to the presence of a ``light ring''.
For example, a classical Schwarzschild black holes has a region at radius $3 r_{s}/2$  where there is an unstable closed orbit for photons.
These orbits define a real (orbital) frequency, plus an imaginary part that corresponds to the decay rate for the photon to fall into or away from the hole.
For classical holes, this complex frequency dominates the ring down signal and coincides with good accuracy to the least damped QNM \cite{PhysRevD.79.064016}.
However, if the classical physics is modified close to the horizon but not at the light ring, the QNM spectrum may change drastically without modifying the leading ring down signal from a merger or infalling object \cite{PhysRevLett.116.171101, Mirbabayi:2018mdm}.

Radiation emitted during the earlier phase of the merger will orbit the newly formed hole and be emitted in a way determined by the characteristics of the light ring, rather than the horizon, and this signal would dominate the first part of the ring down phase.  This makes it unclear whether one should regard this first phase of the ring down signal from a Bekenstein-Mukhanov black hole as controlled by \eqref{wfund}. 

If the physics is only modified very close to the horizon so that the light ring is unaffected, there will still be an effect on gravitational wave emission.  Classical black holes absorb all the radiation that falls into their horizons.  By contrast, a Bekenstein-Mukhanov hole cannot absorb any radiation unless it corresponds to one of the allowed frequencies  \eqref{wfund}.  Waves of other frequencies will be reflected from (or transmitted through) the horizon, and can emerge from the light-ring later when they reach it, possibly after further reflections.  For a modification confined to a distance $\epsilon$ from the classical horizon, these ``echoes'' emerge in a time that scales as $r_{s} \ln r_{s}/ \epsilon$, which is $\sim 16$ times longer than the damping time associated to the light right for a $60$ solar mass black hole and $\epsilon \sim l_{P}$ \cite{PhysRevLett.116.171101, PhysRevD.94.084031}.

Therefore, we distinguish two scenarios:
\begin{itemize}
\item Scenario 1: The Bekenstein-Mukhanov quantization affects the the light ring, so that it can emit and absorb only the frequencies  \eqref{wfund}.  In this scenario, the first part of the ring down signal will differ from the classical prediction and can be used to relatively easily test the proposal.
\item  Scenario 2: The Bekenstein-Mukhanov black hole differs from a classical black hole only in the near-horizon region behind the light ring (concievably only at distances  approximately a Planck length from the horizon).  The physics of the light ring is essentially identical to that of a classical black hole. In this scenario, ``echoes'' that follow the initial ring down may be the only signal of the modification.
\end{itemize}

\subsection{Scenario 1: the light ring can only emit and absorb the frequencies \eqref{wfund}}

The  perturbation  to a classical black hole after a binary merger is mainly  the $l=2$ spherical harmonic, and according to GR the emitted radiation is well-described by the $l=2, k=1$ QNM \cite{Yunes:2016jcc} (the modes at higher $k$ have large imaginary parts and damp rapidly), which coincides with the orbital frequency and decay at the light ring.
The complex frequency of this mode has an imaginary part roughly $1/4$ times its real part \cite{0264-9381-26-16-163001}. The average frequencies of the constituent gravitons emitted in this mode correspond to the real part, with a typical deviation of order the imaginary part.  

By contrast, a Bekenstein-Mukhanov black hole can emit gravitons only with the frequencies determined by \eqref{wfund}, with a width much smaller than the fundamental frequency \cite{Mukhanov:1986me}. Hence one could assume the QNM radiation is well approximated by the Bekenstein-Mukhanov proposal if $\alpha$ is chosen so that \eqref{wfund} (with $n=1$ corresponding to the fundamental transition, and  $m=2$ for a graviton) matches the real part of the $l=2, k=1$ QNM frequency. However, because the QNM frequency is $a$-dependent, this fit would require an $a$-dependent $\alpha$ (Figure \ref{fig:QNMvsBM2}). If $\alpha$ is a universal constant that does not depend on $a$, one observation of a past-merger black hole ringdown with definite $a$ could be used to fix $\alpha$.

%, and then one additional observation with a different $a$ that is consistent with the GR prediction would rule out the theory. 

%\red{I'm not sure anymore, let's talk here.} (Note that this does not actually require any precision in the measurement of $a$, but merely sufficient precision in the measurement of $M$ and $\omega$.)

Of course, if the Bekenstein-Mukhanov proposal is correct there is no guarantee that the precise mass and spin of the final hole inferred  from a fit to classical GR is correct. 
%On the other hand it is improbable that the proposal would predict radiation that agrees with GR for some other set of parameters than the physical ones.  
A  conservative analysis of this proposal would estimate the mass and spin from the inspiral phases alone, where corrections to the black hole geometry according to the Bekenstein-Mukhanov proposal should be small.   Instead, the LIGO analysis uses the entire signal, but most of the statistical power comes from the inspiral and merger phases.  This is because in the events reported thus far, the ringdown phase had  signal-noise of at most about one (see Fig.\ 4 of \cite{TheLIGOScientific:2016src}).  If   the Bekenstein-Mukhanov proposal is correct, the fact that the signal from the inspiral and merger phases appear consistent with GR for \emph{some} set of parameters makes it likely that the  Bekenstein-Mukhanov proposal in fact predicts only small corrections to GR for these phases, so that indeed the GR-based estimates of the mass and spin will not be far off.  Therefore we will focus on the ring-down phase for the purpose of trying to test the  Bekenstein-Mukhanov proposal.  

\begin{figure}
	\includegraphics{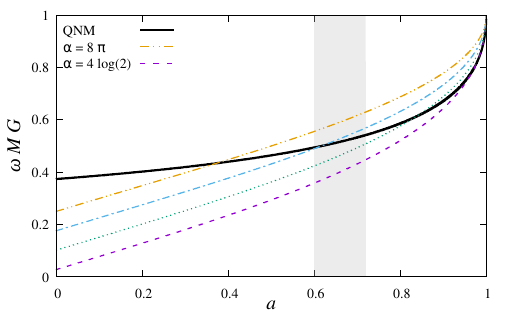}
	\caption{\label{fig:QNMvsBM1}The dimensionless quantity $\omega M G$, where $\omega$ is the real part of the $l=2, k=1$ QNM frequency (thick solid line, using data from \cite{0264-9381-26-16-163001}), and  the fundamental transition frequency $\omega_{n=1, m=2}$ of the Bekenstein-Mukhanov model (\ref{wfund}) for a range of values of $\alpha$ (dashed  lines, $\alpha$ from $4 \ln 2$ to $8 \pi$ in equal steps), as a function of dimensionless spin parameter of the hole $a \equiv J/G M^{2}$.  For reference, the  90\% confidence interval on the  measurement of $a$ for final hole in the LIGO event GW150914 \cite{Abbott:2016blz} is shown (vertical band).  We do not illustrate the uncertainty in $\omega M$, because the ringdown was not detected in this event.   }
\end{figure}

\begin{figure}
	\includegraphics{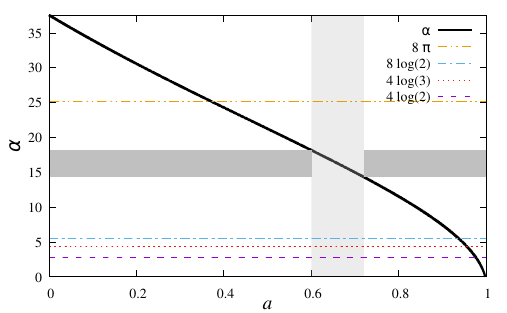}
	\caption{\label{fig:QNMvsBM2}The value of $\alpha$ for which the fundamental transition frequency of the Bekenstein-Mukhanov model matches  the real part of the $l=2, k=1$ QNM frequency, as a function of the black hole's spin $a$.  For reference, the  90\% confidence interval on the  measurement of $a$ for final hole in the LIGO event GW150914 is shown (vertical band) along with the corresponding uncertainty in $\alpha$ (horizontal band).  A single observation of a ring down consistent with GR for a hole with the same (or similar) $a$ as GW150914 would rule out all the extant proposals for $\alpha$.}
\end{figure}

Because the main ringdown signal is controlled by the physics of the light ring, that part of the signal will be strongly modified under Scenario 1.  For instance, suppose the ringdown signal is observed to have a frequency such that $\omega M < 0.37$.  That would be incompatible with GR regardless of the spin $a$, but compatible with the Bekenstein-Mukhanov proposal for many values of $\alpha, a$ (see Fig.~\ref{fig:QNMvsBM1}).  The same is true for some values of $\omega M > 0.37$ that are incompatible with the GR estimate for $a$.

The upcoming LIGO signals will have increased signal to noise.  This should make a detection of the primary ringdown frequency possible, which as described above provides a strong test.  If  at least one harmonic can be detected the test becomes even stronger.  No matter the value of the parameters $\alpha$ and $M$ and $a$ for the remnant, the spectrum of QNM frequencies is incompatible with $\eqref{wfund}$. Measuring the frequency of a higher harmonic is possible with Advanced LIGO and Virgo at design sensitivity, given a sufficiently large number of observations \cite{Brito:2018rfr}.

One can infer the so-called ``chirp mass" $\mathcal{M} \equiv (M_1 M_2)^{3/5} (M_1 + M_2)^{-1/5}$ to high accuracy from the inspiral phase using only Newtonian dynamics and the quadrupole formula for gravitational radiation, and the total mass can be inferred from the last phase of the inspiral (using GR). 
%\red{(we need to say something here, because the merger-phase should also be modified with bek-muk.)}. 
Including a third  detector (such as Virgo), the projected accuracies for the mass parameters are in  are 15-20\% range, and around 5\% for the spin $a$ \cite{PhysRevD.94.104070}.
For a merger of two approximately equal mass holes,  the dimensionless spin of the final hole should be very roughly $a \sim 0.5$. This is because nearly all the orbital angular momentum of the final phase of the inspiral ends up as spin. The orbital angular momentum of a test particle of mass $M_{1}$ at the innermost stable orbit of a Schwarzschild black hole of mass $M_{2}$ is $J = 2 \sqrt{3} G M_{1}  M_{2}$. Linear extrapolation from the extreme mass ratio limit gives a crude estimate of the spin of the final hole after a binary merger of non-spinning holes: $ a \equiv J/G (M_{1}+M_{2})^2 \approx  2 \sqrt{3} \nu$, where $\nu \equiv M_{1} M_{2}/(M_{1} + M_{2})^{2}$.  This is too large; a fraction of the angular momentum is radiated by gravitational waves during the the last phase before the merger \cite{PhysRevD.76.064034}.  A more accurate approximation (based on a fit to numerical simulations) for a binary merger of non-spinning holes is $a \approx  2 \sqrt{3} \nu - 3.87 \nu^{2} \approx 0.62$ when $M_{1} = M_{2}$ \cite{Hofmann:2016yih}. Taking into account the varying mass and spin of the initial holes -- which can range from aligned to anti-aligned with the orbital angular momentum -- gives a fairly large scatter around this value.  The point here is that a reasonable guess for the likely range for the spin of the final hole in an equal mass merger is perhaps $0.4 \lesssim a \lesssim 0.9$. (see \emph{e.g.} Figure 8 of \cite{Kesden:2010yp}).

In Figure \ref{fig:QNMvsBM2} we illustrate the uncertainty on the spin of the final hole in the LIGO event GW150914 \cite{Abbott:2016blz}, and  what would be the corresponding  uncertainty in $\alpha$.   Because the signal/noise of this event was not sufficient to measure the ring down  this event cannot actually be used to constrain $\alpha$.  Nevertheless it provides a rough guide in understanding what the precision of the constraint will be.  The firm expectation of LIGO  is that future events will have higher signal/noise and allow for a measurement of the ring down \cite{Brito:2018rfr,Yang:2017zxs}. Presumably, these future events will also allow for  improved precision on $a$.  In any case, the 90\% confidence interval in $a$ for GW150914 was already small enough that the corresponding uncertainty in $\alpha$ is $14 < \alpha < 18$.  Therefore, an observation of two holes with significantly different $a$ would then provide a strong test of the Bekenstein-Mukhanov proposal.

%\red{It is worth mentioning that the current advanced LIGO parameter estimation for the final mass and spin of the remnant black hole is utilizing all three stages of the waveform. A consistent analysis within this proposal requires the estimation of mass and spin from the inspiral alone, where corrections to the black hole geometry according to the Bekenstein-Mukhanov proposal are still small. However, the ringdown in the case of GW150914 had a signal to noise ratio of about one, so the ringdown signal did not constrain the parameter estimation significantly, see fig.\ 4 of \cite{TheLIGOScientific:2016src}.}

\subsection{Scenario 2: the light ring is unaffected, but the horizon can only emit and absorb the ``allowed'' frequencies \eqref{wfund}}

This scenario is more difficult to conclusively rule out, as it is at least logically possible that the pre-merger and early-time post-merger signal could be identical to classical GR.  This is because the first part of the ringdown signal is determined by the physics of the lightring, not the physics of the horizon.
However, at some time after the merger, the black hole surface would reflect almost all the incoming radiation, except the components with  frequency \eqref{wfund}.

In linear perturbation theory, the ringdown signal will  be described  by the wave equation
\begin{align}
\left[-\partial_t^2 + \partial_x^2 - V(x) \right] \Psi(t,x) = 0
\end{align}
in terms of the tortoise coordinate $x(r) = r + 2 M G \log \left(\frac{r}{2 M G}- 1\right)$. We assume that the black hole is modified near the horizon at $r_0 \leq 2 M G + \epsilon$, for some  $\epsilon < M G$.  If the modification is confined to a Planck length from the horizon, $\epsilon \sim l_P$.

As $r \to \infty$,
\begin{align}
	\Psi(t,x\rightarrow \infty) \propto e^{i \omega (x-t)}.
\end{align}
At $r=r_0$ we use the modified boundary conditions
\begin{align}
	\Psi(t,x_0) \propto e^{i \omega (x_0+t)} + R(\omega) e^{i \omega (x_0-t)},
\end{align}
where $R$ is the reflection coefficient.   Setting $R=0$ reproduces the standard GR calculation of the QNMs.  However according to Bekenstein-Mukhanov, $R(\omega)$ is close to one except at the special values $\eqref{wfund}$. The $r=r_0$ surface effectively acts as a band-stop filter for the reflected radiation.

This  allows long lived modes trapped between the approximately reflective surface and the area close to the light ring. These arise from gravitational waves that fall inward towards $r_0$, reflect because their frequency does not correspond to \eqref{wfund}, and are subsequently rescattered from the light ring and escape eventually to infinity \cite{PhysRevD.94.084031}.

A smoking gun for these modified boundary conditions at $r=r_0$ would be the detection of ``echoes'' and long lived wave trains.
These late-time signals will have a characteristic shape given that they are ``filtered'' by the frequency-dependent absorption of the horizon. Even if the modification is confined to a distance of order one Planck length from the horizon $\epsilon \approx l_P$, the echoes can still be observable -- the effect of decreasing $\epsilon$ is to delay the echoes by a time $\Delta t \approx - \beta M \log \left(\frac{\epsilon}{M G}\right), \ell \ll M$ that scales only logarithmically in the distance to the horizon (where $\beta$ is a factor of order one). For a solar mass black hole with $\epsilon = l_P$ we have $\Delta t \approx 10^{-3} s$.  There is an ongoing discussion of gravitational wave echoes in the literature (see \cite{Abedi:2016hgu,Westerweck:2017hus} and followups).

The echoes predicted by the Bekenstein-Mukhanov proposal would be highly characteristic, because the  the band-stop filter boundary conditions imprint the spectrum $\eqref{wfund}$ on the outgoing radiation.
As a long-standing proposal with a relatively clear theoretical basis, horizon area quantization may provide a stronger motivation to search for such echoes than ad-hoc modifications of the near-horizon geometry (such as {simply} replacing the horizon with a reflective boundary condition).

\section{Conclusions}
We have proposed two scenarios under which the Bekenstein-Mukhanov proposal for the quantization of black hole horizon area can be tested by future data from black hole mergers. In upcoming work \cite{cfk}, we will investigate the scenario where only the horizon is strongly modified in more detail.

We conclude by commenting on some possible issues and open questions.
\begin{itemize}
	\item The primary uncertainty is the merger phase itself.  It is very plausible that the physics  deviates strongly from  classical GR, making the Bekenstein-Mukhanov proposal easy to rule out (or confirm).  Unfortunately, lacking a detailed model of the dynamics we cannot make use of this.
	
	\item In Scenario 2,  the light ring is unmodified and one must rely on ``echoes'' that are emitted after the initial ring down.  It is not clear yet what the amplitude of this signal would be, or what form it would take. Possibly it is too weak to detect.
	
	\item If $\alpha$ depends on $a$, the fundamental frequency could match the $l=2, n=1$ QNM frequency for all $a$ (Figure \ref{fig:QNMvsBM2}).  However, this would mean there is no fundamental quantum of area, which is inconsistent with the motivation for the original proposal.  Furthermore, even this possibility could be ruled out if the $k \neq 1$ or $l \neq 2$ QNMs were observed, since their frequencies  do not correspond to \eqref{wfund} with $n>1$.
	
	\item As mentioned above, if more than one graviton is emitted per area transition, the frequency of each is continuously variable.  However, unless $\alpha$ is much larger than any of the proposals in the literature, these {gravitons} would have frequencies well below those predicted by GR.  Furthermore there is no reason single graviton emission would be subdominant -- instead, as Bekenstein argues in \cite{Bekenstein:1997bt}, single graviton emission should dominate -- and in any case one would have to argue why this sort of emission would match the  predictions of GR.
	
	\item We assumed the dominant area transition is the  minimal one: $n = -\Delta N =1$.  If  $n \neq 1$, one must have a single $n$ that dominates the spectrum, or else multiple harmonics would be observed (which would not be consistent with GR).  Larger $n$ then just renormalizes $\alpha \to n \alpha$, and can be ruled out in the same way as $n=1$.   If $n$ is a function of $a$, there would have to be some values of $a$ where more than one harmonic is visible, conflicting with GR. 
		
	\item Finally, it is possible that the huge degeneracy at each area level is in fact split into a near continuum without gaps, with the spacing everywhere in the spectrum being $\sim e^{-S}$. This would be impossible to rule out, but would mean for macroscopic holes the area is quantized in much, much, much finer units than the Planck area, invalidating the Bekenstein-Mukhanov proposal.
\end{itemize}

\begin{acknowledgments}
The work of MK is supported in part by the NSF through grant PHY-1214302, and he
acknowledges membership at the NYU-ECNU Joint Physics Research Institute in
Shanghai. We thank Savas Dimopoulos for the discussions that initiated this work, as well as Vitor Cardoso, Gia Dvali, Andrei Gruzinov, Slava Mukhanov, Alberto Nicolis, Massimo Porrati, Kris Sigurdson, and the  anonymous referees for useful comments.
\end{acknowledgments}

\bibliography{ref}

\end{document}